\documentclass[
prc,%
10pt,%
final,%
notitlepage,%
oneside,%
twocolumn,%
nobibnotes,%
nofootinbib,% In the current version of REVTeX, when this option is on,
superscriptaddress,%
floatfix]%
{revtex4}
\usepackage{color} %{\color{red} }
\usepackage{amsfonts} 
\usepackage{amsbsy} 
\usepackage{mathrsfs}
\usepackage{graphicx}
\usepackage{comment}

\def\lsim{\mathrel{\rlap{ \lower4pt\hbox{\hskip-3pt$\sim$}}
    \raise1pt\hbox{$<$}}} %less than approx. symbol
\def\gsim{\mathrel{\rlap{ \lower4pt\hbox{\hskip-3pt$\sim$}}
    \raise1pt\hbox{$>$}}} %greater than or approx. symbol
\def\scr#1{\mbox{\scriptsize #1}}

\begin{document}
% ============================================================

\title{Predictions of baryon directed flow in heavy-ion collisions at high baryon density} 
%\title{Proton Antiflow at $\sqrt{s_{NN}}=$ 7.2 GeV} 

%
\author{Yuri B. Ivanov}\thanks{e-mail: yivanov@theor.jinr.ru}
\affiliation{Bogoliubov Laboratory of Theoretical Physics, Joint Institute for Nuclear Research,
  141980 Dubna, Russia} 
\affiliation{National Research Center
  "Kurchatov Institute", 123182 Moscow, Russia}
%
%\author{M. Kozhevnikova}\thanks{e-mail: kozhevnikova@jinr.ru}
%\affiliation{Veksler and Baldin Laboratory of High Energy Physics,
%  JINR Dubna, 141980 Dubna, Russia}

\begin{abstract}
Predictions of the proton directed flow ($v_1$) in semicentral Au+Au collisions 
in the energy range between 4.5 and 7.7 GeV are done. 
The calculations are performed within the model of three-fluid dynamics  
with crossover equation of state, which well reproduces the proton $v_1$ both below 4.5 GeV and above 7.7 GeV, 
as well as bulk observables in the energy range of interest.  
It is predicted that the proton flow evolves non-monotonously. 
At the energy of 7.2 GeV it exhibits antiflow (i.e. negative slope of 
$v_1(y)$) in the midrapidity. At 7.7 GeV, the flow returns to the normal pattern in accordance with the 
STAR data. 
The midrapidity $v_1$-slope excitation functions within the first-order phase and 
crossover transitions to quark-gluon phase (QGP) turn out to be 
qualitatively similar, but the amplitude of the wiggle in the crossover scenario is much smaller than 
that in the strong first-order phase transition. Therefore, 
the change of sign followed by minimum at 7.2 GeV in the $v_1$-slope excitation function 
indicates onset of (weak phase or crossover) transition to QGP. 
The second change of the sign around 10 GeV 
results from interplay between incomplete baryon stopping and transverse expansion of the system.  
%that was predicted by R.~J.~M.~Snellings {\it et al.} [Phys. Rev. Lett. \textbf{84}, 2803 (2000)]. 
%
%  \pacs{25.75.-q, 25.75.Nq, 24.10.Nz} 
%	\keywords{relativistic heavy-ion collisions, hydrodynamics, directed flow}
\end{abstract}
\maketitle
% \today

% ______________________________________________________________________
\section{Introduction}

The directed flow is defined as the first coefficient, $v_1$, in the Fourier expansion of particle distribution, 
$d^2N/(dy\; d\phi)$, in azimuthal angle $\phi$ with respect to the reaction plane \cite{Voloshin:1994mz,Voloshin:2008dg}: 
\begin{eqnarray}
 \label{vn-def}
\frac{d^2 N}{d y\;d\phi} = \frac{d N}{dy}
\left(1+ \sum_{n=1}^{\infty} 2\; v_n(y) \cos(n\phi)\right),
\end{eqnarray}
where $y$ is the longitudinal rapidity of a particle. Equation (\ref{vn-def}) defines so-called transverse-momentum integrated flow coefficients, which are considered below. 

The discussion below primarily relates to heavy-ion collisions energies 
$\sqrt{s_{NN}}=$ 3--12 GeV, at which high baryon density is achieved. 
This energy range is actively explored in the Beam-Energy
Scan (BES) program at RHIC \cite{Aparin:2023fml} and NA61/SHINE at
SPS \cite{shine}, and will be further studied in forthcoming facilities \cite{Galatyuk:2019lcf}: 
NICA \cite{nica}, FAIR \cite{fair}, HIAF \cite{HIAF}, and J-PARC-HI \cite{jparc-hi}.

The directed flow is one of the most sensitive quantities to the equation of state (EoS) of strongly-interacting matter. 
Moreover, it provides signals of phase transition to the quark-gluon phase (QGP) 
\cite{Hung:1994eq,Rischke:1995pe,Rischke:1996nq,CR99,Br00,St05}. However, 
the directed flow depends not only on the EoS. First of all, it strongly depends on the stopping power of the nuclear matter. 
The finite stopping power can even mimic the phase transition effect \cite{Snellings:1999bt}. 
Though, the stopping power can be constrained proceeding from bulk observables for protons and pions. 
In Ref. \cite{Du:2022yok}, the baryon stopping was even deduced from data on the the directed flow.  
The directed flow of various hadrons is even more involved. 
In heavy-ion collisions at high baryon density, 
the directed flow of pions, anti-kaons, $\Lambda$s, {\it etc.} 
are noticeably modified during the afterburner stage \cite{Ivanov:2024gkn,Batyuk:2016qmb,Batyuk:2017sku}, 
where the nonequilibrium dominates rather then the EoS effects. 
When the time for the nuclei to pass each other becomes long relative to
the characteristic time scale for the participant evolution,
the interaction between participants and spectators (so-called shadowing) becomes important
\cite{Ivanov:2024gkn,Bass:1993ce,Heiselberg:1998es,Liu:1998yc}, which strongly modifies 
the directed flow of various hadrons but not of protons and kaons \cite{Ivanov:2024gkn}.
The proton flow is only slightly modified by the afterburner and shadowing \cite{Ivanov:2024gkn,Batyuk:2016qmb,Batyuk:2017sku} 
because the baryon matter itself shadows escape of other hadrons. 
The kaons have a long mean free path and hence are not affected by this shadowing. On the other hand, 
because of the same long mean free path, the kaons early decouple from the expanding fireball, 
probably even before the fireball becomes thermalized.  
Summarizing these arguments, the baryon (proton) directed flow appears to be the most promising observable 
for studying the EoS effects in heavy-ion collisions at high baryon density.

Data on the directed flow in Au+Au collisions at $\sqrt{s_{NN}}\geq$ 7.7 GeV were reported by the STAR collaboration 
in Ref. \cite{STAR:2014clz}. The analysis of these data was performed within various approaches 
\cite{Du:2022yok,Konchakovski:2014gda,Ivanov:2014ioa,Ivanov:2016sqy,Steinheimer:2014pfa,Nara:2016phs,Shen:2020jwv,Ryu:2021lnx,Nara:2016hbg,Nara:2020ztb,Nara:2019qfd,Nara:2021fuu,Nara:2022kbb},
which include both hydrodynamic and kinetic models. These studies indicated that 
the transition to the quark-gluon phase is most probably of the crossover or weak-first-order type \cite{Du:2022yok,Konchakovski:2014gda,Ivanov:2014ioa,Ivanov:2016sqy}. 
It was suggested \cite{Du:2022yok} that the observed sign change in the midrapidity slope of the proton $v_1$ 
around 10-20 GeV collision energy is not an evidence of the first-order phase transition of the QCD
matter, but rather a consequence of the initial baryon stopping as it was predicted in Ref. \cite{Snellings:1999bt}.

The STAR-FXT (fixed-target) data on the directed flow of identified particles at energies 
$\sqrt{s_{NN}}=$ 3 and 4.5 GeV were recently published in Refs. \cite{STAR:2020dav,STAR:2021yiu}. 
These data were also analyzed within various, mostly kinetic models  
\cite{Ivanov:2024gkn,Nara:2020ztb,Nara:2022kbb,Oliinychenko:2022uvy,Steinheimer:2022gqb,OmanaKuttan:2022aml,Li:2022cfd,Wu:2023rui,Parfenov:2022brq,Mamaev:2023yhz,Yao:2023yda,Kozhevnikova:2023mnw,Kozhevnikova:2024itb,Yong:2023uct,Wei:2024yda}  
in relation to the EoS of the matter produced in these collisions. 
The kinetic models discussed the directed flow in terms of softness and stiffness of the EoS 
\cite{Nara:2020ztb,Oliinychenko:2022uvy,Steinheimer:2022gqb,OmanaKuttan:2022aml,Wu:2023rui,Yao:2023yda}. 
They indicated 
%\cite{Nara:2020ztb,Oliinychenko:2022uvy,Steinheimer:2022gqb,OmanaKuttan:2022aml,Wu:2023rui,Yao:2023yda}
preference of stiff (to a different extent) EoSs for the reproduction of the directed flow  
at $\sqrt{s_{NN}}=$ 3 GeV, while the $v_1$ data at 4.5 GeV required a softer EoS. 
The latter was interpreted as indication of onset of transition into QGP. 
The conclusion about preference of the stiff EoS at 3 GeV contradicts earlier findings.
Strong preference of the soft EoS was reported in Refs. 
\cite{Ivanov:2014ioa,Ivanov:2016sqy,Pal:2000yc,Danielewicz:2002pu,Russkikh:2006ae}. 
In Refs. \cite{Ivanov:2014ioa,Ivanov:2016sqy}, the EoS is additionally softened at  $\sqrt{s_{NN}} >$ 4 GeV
because of onset of the deconfinement transition. The same conclusion resulted from the analysis of data at 
%$\sqrt{s_{NN}}=$ 
3 and 4.5 GeV within hydrodynamic model \cite{Ivanov:2024gkn,Kozhevnikova:2023mnw,Kozhevnikova:2024itb}.  
A more extended discussion of the EoS constraints deduced from the directed-flow analysis 
is presented in review \cite{Sorensen:2023zkk}.

Very recently, the data on the directed flow (in particular, the proton one) were presented \cite{STAR:2025twg}
at energies 3.2, 3.5 and 3.9 GeV. These data bridge the STAR results at 3 and 4.5 GeV \cite{STAR:2020dav,STAR:2021yiu}.
However, a very interesting energy range between 4.5 and 7.7 GeV remains experimentally unexplored. Precisely in this 
energy range, the most spectacular signals of onset of the deconfinement transition are expected.

In this paper the new data on the proton directed flow \cite{STAR:2025twg} at energies 3.2, 3.5 and 3.9 GeV are described. 
Also predictions of the proton directed flow in the energy range between 4.5 and 7.7 GeV are done. 
The calculations are performed within the model of three-fluid dynamics (3FD) \cite{Ivanov:2005yw,Ivanov:2013wha}. 
Different EoSs can be implemented in the 3FD model. 
Right now, three different EoSs are used in the 3FD simulations: 
a purely hadronic EoS \cite{gasEOS} and two EoSs with deconfinement transitions \cite{Toneev06}, i.e. 
an EoS with a first-order phase transition (1PT EoS) and one with a smooth crossover transition.  
The crossover EoS well reproduces the proton directed flow both below 4.5 GeV and above 7.7 GeV. 
In addition, it well describes bulk observables \cite{Ivanov:2013yqa,Ivanov:2018vpw} 
in the energy range of interest. Therefore, the crossover predictions for the proton directed flow 
in the energy range from 4.5 to 7.7 GeV can be considered as reliable. Predictions with two other 
EoSs are also presented to illustrate sensitivity of the proton directed flow to the EoS.

%===================================================================
\section{3FD model and THESEUS} 
  \label{3FD and THESEUS}

The 3FD approximation is a minimal way to simulate the early-stage nonequilibrium in the colliding nuclei. 
The 3FD model \cite{Ivanov:2005yw,Ivanov:2013wha} describes nonequilibrium at
the early stage of nuclear collisions by means of two counterstreaming baryon-rich fluids. 
In addition, newly produced particles, populating
predominantly the midrapidity region, are attributed to a third, so-called fireball fluid.
These fluids
%, i.e. the projectile (p), target (t), and fireball (f), 
are governed by conventional hydrodynamic equations coupled by 
friction terms in the right-hand sides of the Euler equations. The friction terms 
describe the energy--momentum exchange between the fluids. 
The 3FD model describes a nuclear collision from the stage of the incident cold nuclei
approaching each other, to the final freeze-out 
%stage. The hydrodynamic evolution ends with the freeze-out procedure described in Refs.
\cite{Russkikh:2006aa,Ivanov:2008zi}.  
The freeze-out criterion is $\varepsilon < \varepsilon_{\scr{frz}}$, 
where $\varepsilon$ is the total energy density of all three fluids in their common rest frame.
The freeze-out energy density $\varepsilon_{\scr{frz}}=0.4~$GeV/fm$^3$ 
was chosen mostly on the condition of the best reproduction 
of secondary particle yields for all considered EoSs, see \cite{Ivanov:2005yw}.

The 3FD model does not include any kinetic afterburner stage. 
The THESEUS event generator 
(Three-fluid Hydrodynamics-based Event Simulator Extended by UrQMD final State interactions)
\cite{Batyuk:2016qmb,Batyuk:2017sku,Kozhevnikova:2020bdb}, which follows the 3FD simulation,
does include the afterburner stage that is described by the UrQMD 
(Ultrarelativistic Quantum Molecular Dynamics) model \cite{Bass:1993ce}.
The afterburner stage is of prime importance for   
collisions at lower energies, where there is no clear rapidity separation between
participants and spectators at the freeze-out. 
When the time
for the nuclei to pass each other becomes long relative to
the characteristic time scale for the participant evolution, 
the expansion of the fireball, consisted of participants, is 
shadowed by spectators \cite{Bass:1993ce,Heiselberg:1998es,Liu:1998yc,Ivanov:2024gkn}.

\begin{figure}[!tbh]
\includegraphics[width=.47\textwidth]{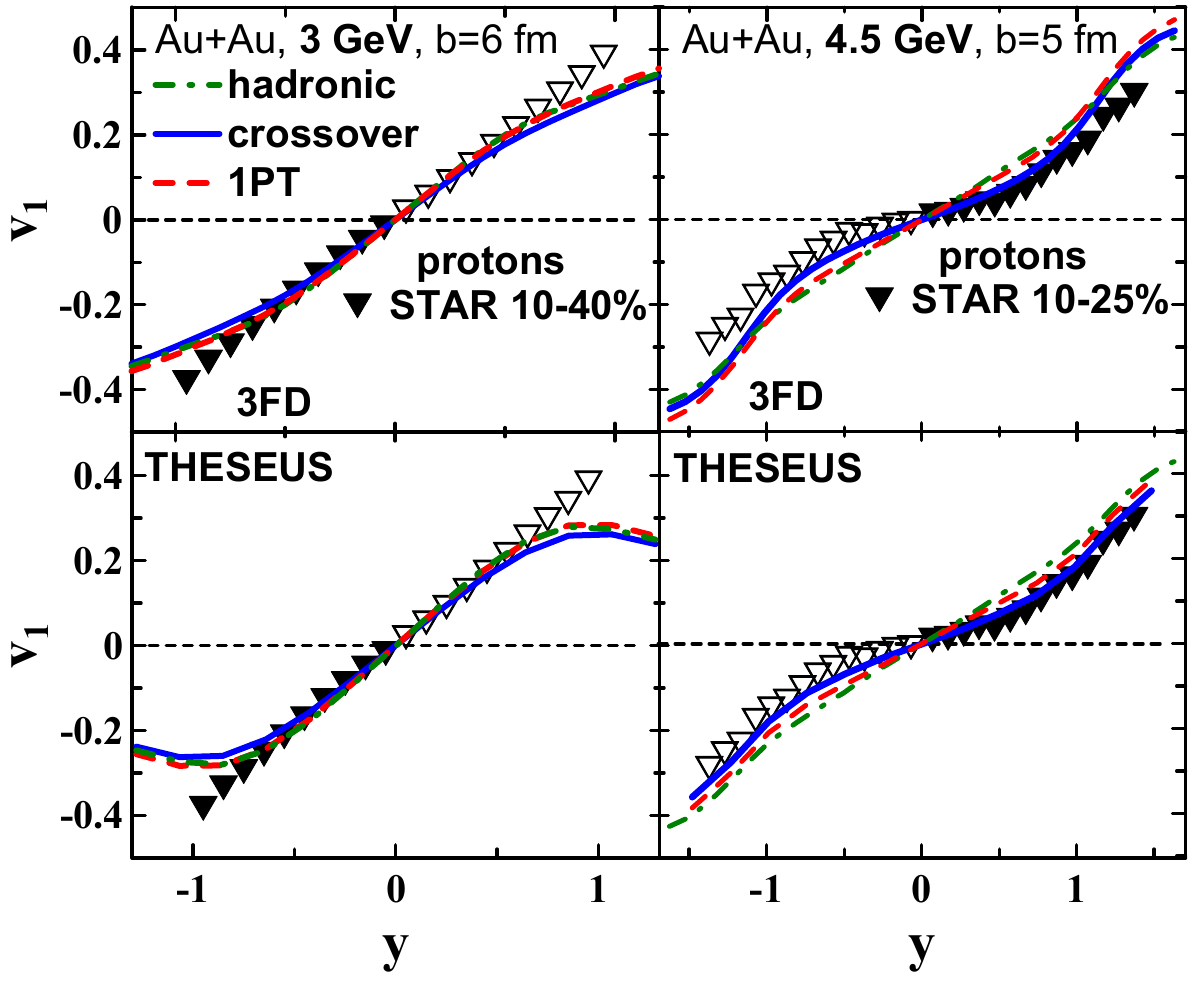}
  \caption{(Color online)
Directed flow of protons  as function of rapidity in semicentral %($b=$ 6 fm)    
Au+Au collisions at collision energies of $\sqrt{s_{NN}}=$ 3 and 4.5 GeV. 
Results are calculated within the 3FD model (upper row of panels) and 
the THESEUS \cite{Ivanov:2024gkn} (lower row of panels) with hadronic, 1PT, and crossover EoSs. 
STAR data are from Refs. \cite{STAR:2020dav,STAR:2021yiu}.
}
    \label{fig:v1-STAR-fxt_3FD-THESEUS}
\end{figure}
\begin{figure*}[!htb]
\includegraphics[width=.87\textwidth]{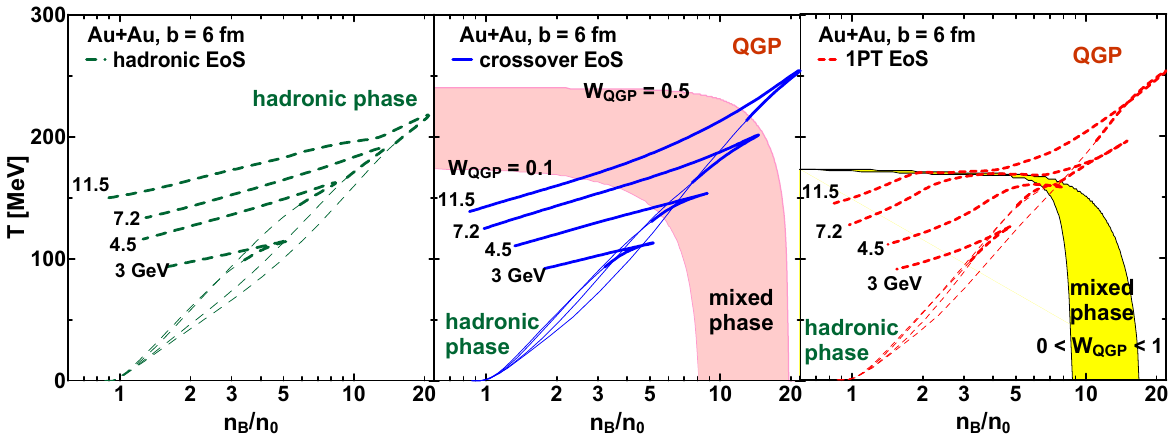}
 \caption{(Color online)
Dynamical trajectories of the matter in the central cell of the colliding Au+Au nuclei  
in semicentral collisions (impact parameter is $b=$ 6 fm) at energies $\sqrt{s_{NN}}=$ 3, 4.5, 7.2 and 11.5 GeV. 
The trajectories are plotted in terms of the baryon density 
($n_B$, scaled by the normal nuclear density $n_0$) and temperature $T$. 
The trajectories are presented for the three EoSs. 
The wide shadowed area displays the region of the crossover EoS, where the QGP fraction $W_{QGP}$ lies between
0.1 and 0.5. The shadowed region in the 1PT panel indicates the mixed phase, where $0<W_{QGP}<1$. 
}
\label{fig:T-nB_b=6fm}
\end{figure*}

Not all observables are equally modified during the afterburner stage.  
The proton directed flow is only slightly changed \cite{Ivanov:2024gkn,Batyuk:2016qmb,Batyuk:2017sku}. 
This is one of the reasons for choosing the proton $v_1$ as a preferable observable to quantify the EoS, 
as it was mentioned in the Introduction. Figure \ref{fig:v1-STAR-fxt_3FD-THESEUS} illustrates 
dependence of the proton $v_1$ on the afterburner. 
As seen, the afterburner does not change the proton $v_1$ in the midrapidity region at 3 GeV and slightly 
changes at forward/backward rapidities. At 4.5 GeV the proton $v_1$ is somewhat modified by the afterburner 
in the midrapidity region making reproduction of the data slightly better. The directed flow of other hadrons 
is noticeably stronger affected by the afterburner \cite{Ivanov:2024gkn,Batyuk:2016qmb,Batyuk:2017sku}.

%%%%%%%%%%%%%%%%%%%%%%%%%%%%%%%%%%%%%%%%%%%%%%%%%%%%%%%%%%%%%%%%%%%%%%%%
\section{Equations of State} 
  \label{EoS}

As has been already mentioned,  
three different EoSs are used in the 3FD simulations: 
a purely hadronic EoS \cite{gasEOS},  
the EoS with a first-order phase transition (1PT EoS) and one with a
smooth crossover transition \cite{Toneev06}.  
The hadronic EoS is quite flexible, it allows for changes of incompressibility. 
The used version of the hadronic EoS is characterized by incompressibility $K=$ 190 MeV. 
All three EoSs are similar in the hadronic phase. 
The crossover pressure starts to deviate from 
the hadronic one at $n_B>$ 4-5$n_0$ at temperatures 100--150 MeV  
that are typical for the collisions at STAR-FXT energies, see Fig. \ref{fig:T-nB_b=6fm}.

Dynamical trajectories of the matter in the central cell of the colliding Au+Au nuclei  
in semicentral collisions ($b=$ 6 fm) at energies $\sqrt{s_{NN}}=$ 3, 4.5, 7.2 and 11.5 GeV
are presented in Fig. \ref{fig:T-nB_b=6fm} in terms of the baryon density 
and temperature.
Evolution starts from the normal nuclear density and zero temperature. 
Thermalization of the matter in this central cell occurs 
shortly before reaching the turning point, at which density and temperature
are maximal \cite{Ivanov:2019gxm}. 
Only after the thermalization the temperature takes its conventional meaning. 
These thermalized parts of the trajectories are displayed by bold lines. 
The trajectories for the hadronic and crossover EoSs differ only quantitively. 
The 1PT trajectories show also qualitative differences. 
The 7.2-GeV and 11.5-GeV trajectories exhibit wiggles related to the turn along the mixed-phase region. 
Even the 4.5-GeV trajectory demonstrates certain flattening after the turning point. 
This behavior is related to the well-known reheating phenomenon in the first-order phase transition. 
The temperature along these wiggles does not rise but rather stays nearly constant because the 
reheating competes with rapid dynamical expansion of the system.

The crossover transition constructed in Ref. \cite{Toneev06} is very smooth, it is seen from Fig. \ref{fig:T-nB_b=6fm}.  
Similarly smooth crossover is also implemented in the PHSD model (Parton-Hadron-String Dynamics) \cite{Cassing:2009vt}. 
Such a smooth crossover \cite{Toneev06} certainly contradicts the
lattice QCD data at zero chemical potential \cite{Aoki:2006we}, which indicate a fast crossover.
However, this  shortcoming is not severe for the present simulations, in which the system evolves 
in the region of high baryon densities, where the EoS is not known from the first principles.

When traversing the mixed-phase region of the 1PT EoS, the trajectories pass through the softest-point region,  
where the isentropic speed of sound ($c_s$), which is defined as derivative of 
the pressure ($P$) over the energy density ($\varepsilon$) at constant entropy ($S$)
\begin{eqnarray}
\label{Cs-def}
c_s^2 = \left(\frac{\partial P}{\partial \varepsilon}\right)_{S},  
\end{eqnarray}
exhibits a minimum \cite{Hung:1994eq,CR99,Br00,Toneev:2003sm}.  
At high baryon densities, it is appropriate to speak about the softest-point region rather than the softest point, 
because the EoS is soft in certain region of temperatures and baryon densities \cite{Toneev:2003sm}.
In the softest-point region, the dynamics of the matter slows down 
leading to a longer lifetime of the excited system \cite{Hung:1994eq}. 
This results in a significant reduction of transverse expansion (as compared to what it would be without phase transition)
and, in particular,  to reduction of the directed flow \cite{CR99,Br00}.  
The softness of the EOS affects not only the transverse, but also the
longitudinal expansion.  
It affects the midrapidity region of the 
rapidity distribution of net protons in central collisions  and manifests itself 
as wriggle in the excitation function of the midrapidity curvature of net-proton rapidity distribution 
\cite{Ivanov:2013wha,Ivanov:2012bh,Ivanov:2015vna}.

\begin{figure}[!hbt]
\includegraphics[width=7.6cm]{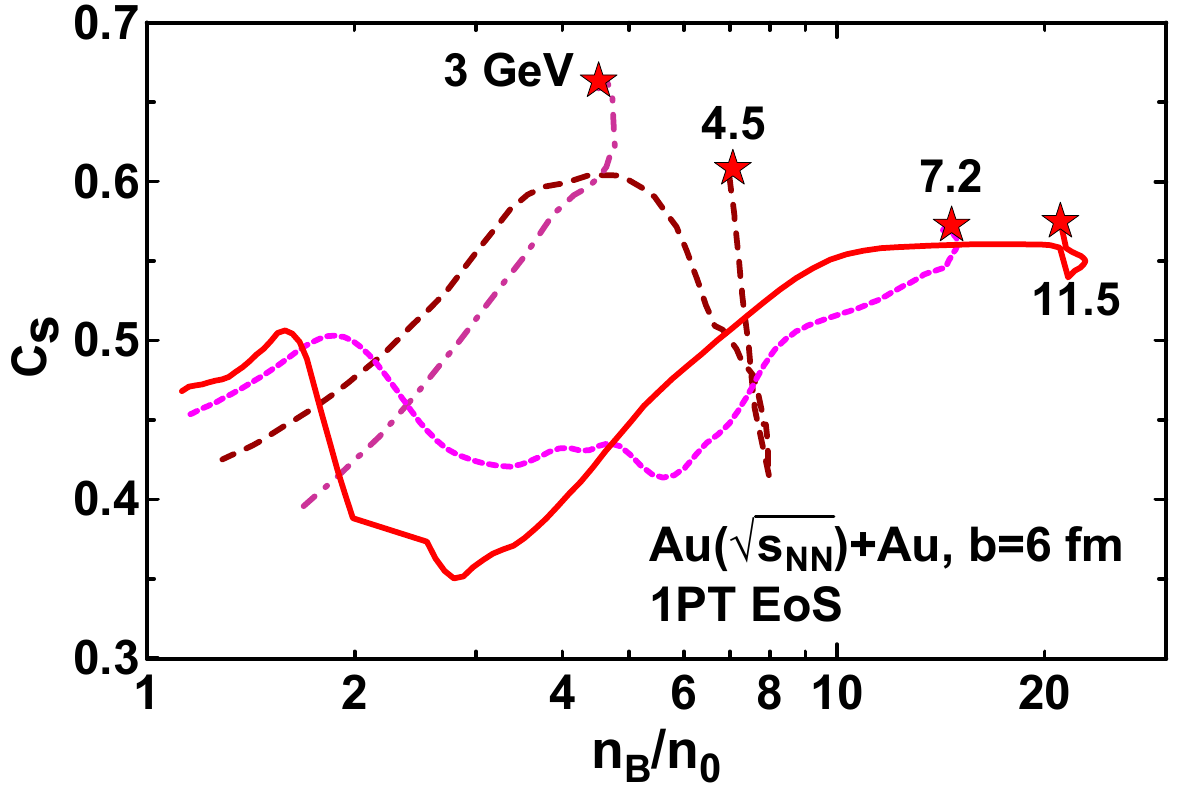}
 \caption{(Color online)
Evolution of the isentropic speed of sound ($c_s$) as function of the 
baryon density ($n_B$, scaled by the normal nuclear density $n_0$) along 
the dynamical trajectories displayed in Fig. \ref{fig:T-nB_b=6fm}. 
The evolution is displayed from the instants (indicated by star symbols) 
that are close to the trajectory turning points, when the matter is sufficiently thermalized. 
The trajectories are presented for the 1PT EoS. 
}
\label{fig:Cs-vs-nb-tph-b6}
\end{figure}
\begin{figure}[!t]
\includegraphics[width=7.6cm]{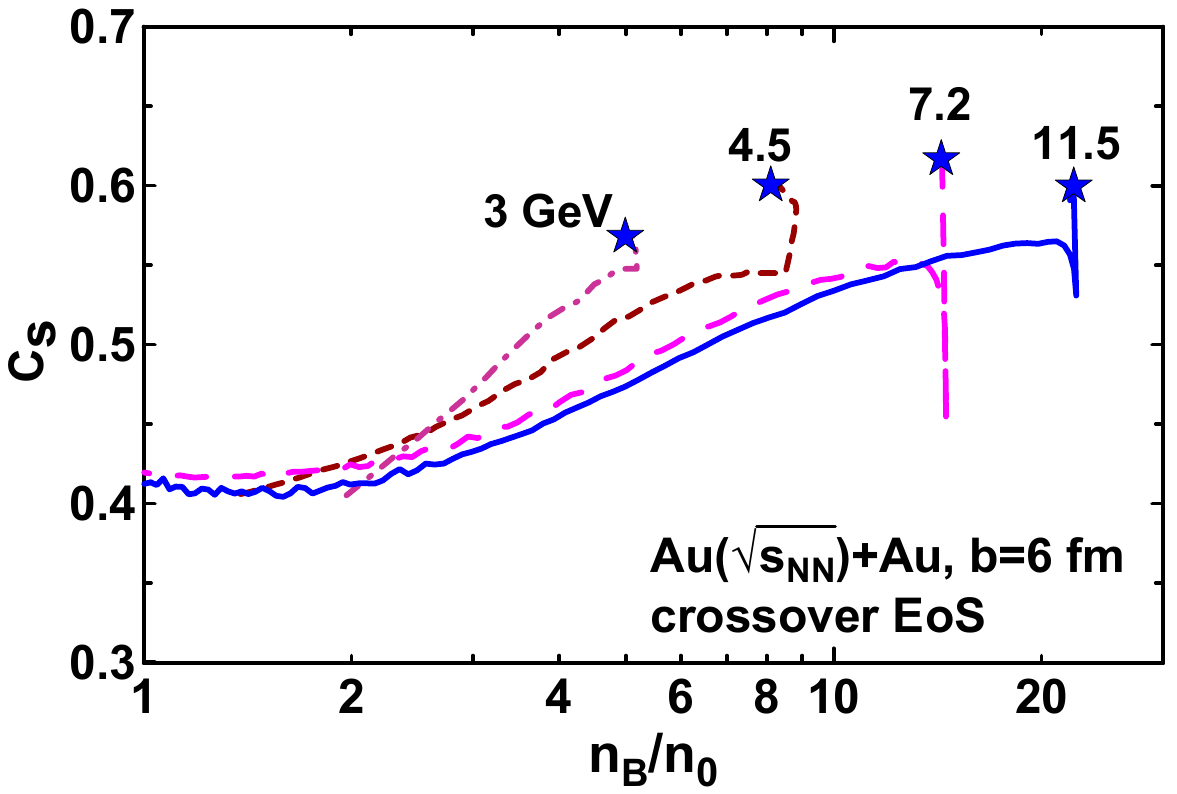}
 \caption{(Color online)
The same as in Fig. \ref{fig:Cs-vs-nb-tph-b6} but for the crossover EoS. 
}
\label{fig:Cs-vs-nb-mix-b6}
\end{figure}

The speed of sound along the trajectories (presented in Fig. \ref{fig:T-nB_b=6fm}) 
is displayed in Figs. \ref{fig:Cs-vs-nb-tph-b6} and \ref{fig:Cs-vs-nb-mix-b6} for the 1PT and crossover EoSs. 
The evolution in Figs. \ref{fig:Cs-vs-nb-tph-b6} and \ref{fig:Cs-vs-nb-mix-b6}  
is displayed beginning from instants (indicated by star symbols) 
when the matter is sufficiently equilibrated \cite{Ivanov:2019gxm} and therefore the thermodynamic definition 
of the  speed of sound is appropriate. The equilibration in the central region is attained 
shortly before reaching the turning point \cite{Ivanov:2019gxm}, at which density and temperature
are maximal, see Fig. \ref{fig:T-nB_b=6fm}. After that the evolution of the unified fluid is 
approximately (up to viscous-like dissipation) isentropic \cite{Ivanov:2016hes} and therefore 
the speed of sound along the trajectory takes the meaning of the isentropic speed of sound.

As seen from Figs. \ref{fig:T-nB_b=6fm} and \ref{fig:Cs-vs-nb-tph-b6}, the turning point and thus the 
softest point of the 4.5-GeV trajectory for the 1PT EoS occurs in the mixed phase region. 
However, this softest-point does not affect the directed flow, 
as will be seen below, because only the central region of the entire system falls within this softest-point region 
and only for a short time. The strong effect on $v_1$ occurs 
at higher collision energies, when a larger part of the matter  
falls within this softest-point region and for a longer time.   
Indeed at energies 7.2 and 11.5 GeV,  
the trajectories for the 1PT EoS turn out to be captured within the mixed-phase region for some time, 
see Fig. \ref{fig:T-nB_b=6fm}. This results in the softest-point regions in the 7.2-GeV and 11.5-GeV trajectories
in Fig. \ref{fig:Cs-vs-nb-tph-b6}. 

The $c_s$ evolution for the crossover EoS is quite monotonous except for the cases of the energies of 7.2 11.5 GeV, 
see  Fig. \ref{fig:Cs-vs-nb-mix-b6}. 
Note that the softest points still exist in the crossover EoS \cite{Toneev:2003sm},  
although they are much less pronounced as compared to the 1PT EoS. 
The crossover 7.2-GeV and 11.5-GeV trajectory exhibit, to different extent, behavior similar to that of 
the 1PT 4.5-GeV one, i.e. exhibit the softest points. 
This has consequences for the proton $v_1$, as it will be seen below.

\section{Directed Flow} 
  \label{Directed Flow}

\begin{figure*}[!hbt]
\includegraphics[width=.98\textwidth]{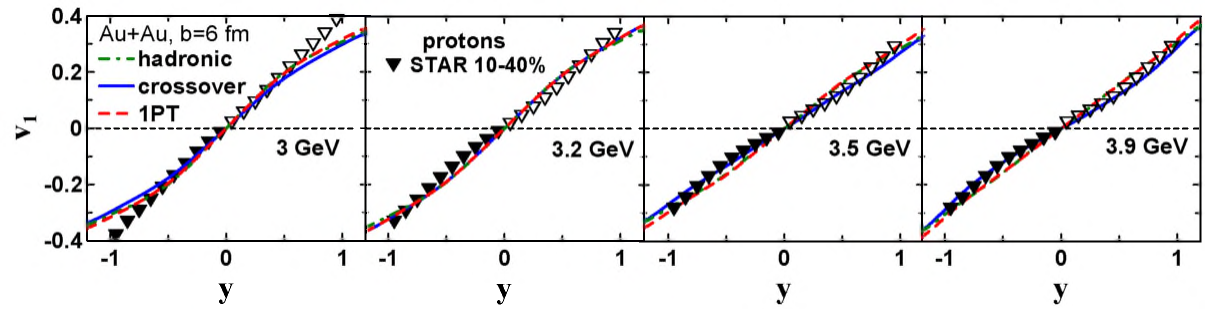}
  \caption{(Color online)
Directed flow of  protons as function of rapidity in semicentral  ($b=$ 6 fm)    
Au+Au collisions at collision energies of $\sqrt{s_{NN}}=$ 3, 3.2, 3.5 and 3.9 GeV. 
Results are calculated within the 3FD model with hadronic, 1PT, and crossover EoSs. 
STAR data are from Refs. \cite{STAR:2021yiu,STAR:2025twg}.
}
    \label{fig:v1-STAR-fxt-3-3,9GeV}
\end{figure*}
\begin{figure*}[!hbt]
\includegraphics[width=.75\textwidth]{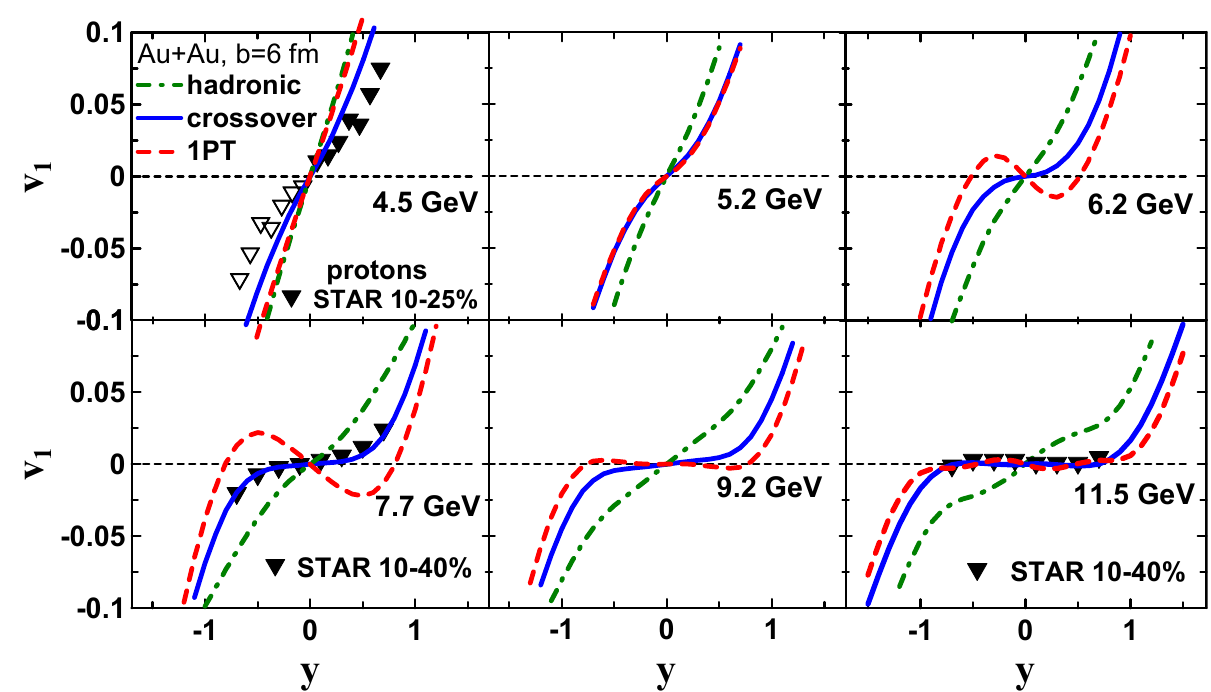}
  \caption{(Color online)
The same as in Fig. \ref{fig:v1-STAR-fxt-3-3,9GeV} but for 
collision energies of $\sqrt{s_{NN}}=$ 4.5, 5.2, 6.2, 7.7, and 11.5 GeV. 
Available STAR data are from Refs. \cite{STAR:2020dav,STAR:2014clz}.
}
    \label{fig:v1-STAR-fxt-4-12GeV}
\end{figure*}

The calculated directed flow of  protons 
as function of rapidity in semicentral     
Au+Au collisions at collision energies of $\sqrt{s_{NN}}=$ 3, 3.2, 3.5 and 3.9 GeV   
is presented in Fig. \ref{fig:v1-STAR-fxt-3-3,9GeV}.
These calculations were performed in the 3FD model with hadronic, 1PT, and crossover EoSs.  
The collision centrality is associated with the corresponding mean impact parameter ($b=$ 6 fm)
by means of the Glauber simulations based on the nuclear overlap calculator\footnote{ 
In all Glauber simulations \cite{web-docs.gsi.de}, including energies $\sqrt{s_{NN}}>$ 4 GeV, the options ``sharp sphere''
and $\sigma_{NN}=30$ mb were used. The results are practically the same if $\sigma_{NN}=40$ mb is taken. 
}
\cite{web-docs.gsi.de}, 
taking into account that colliding nuclei in the 3FD model have the 
sharp-edge (rather than diffuse edge) density profile. 
The results are compared with STAR data \cite{STAR:2021yiu,STAR:2025twg}.

As seen, the proton $v_1$ flow is identical within three considered scenarios at collision energies of 3 and 3.2 GeV.  
All these scenarios equally well reproduce the experimental data. At 3.5 and 3.9 GeV, the crossover scenario
becomes slightly preferable in the midrapidity region. Although its difference from predictions of other 
scenarios remains insignificant. 

Figure \ref{fig:v1-STAR-fxt-4-12GeV} demonstrates the flow at collision energies 4.5--11.5 GeV,
including predictions for the energies between 4.5 and 7.7 GeV. 
The directed flow at 5.2--7.2 GeV  is calculated for $b=$ 6 fm and $p_T$ acceptance of 0.4$ <p_T< $2 GeV/c 
in order it can be directly compared with results at lower \cite{STAR:2021yiu,STAR:2025twg,STAR:2020dav}
and higher \cite{STAR:2014clz} collision energies. 
Note that the directed flow at 4.5 GeV is also presented for $b=$ 6 fm   
so that it can be compared consistently with $v_1$ at other energies. 
While the centrality selection of the data \cite{STAR:2020dav} 
better corresponds to $b=$ 5 fm. 
The Glauber calculator \cite{web-docs.gsi.de} gives the $b$ range between 3.9 and 6.2 fm for 
the centrality interval 10--25\%. The $b$ value is taken just in the middle of this range. 
This is why the reproduction of the data at 4.5 GeV suffers.  
If the proper $b=$ 5 fm is used as in Fig. \ref{fig:v1-STAR-fxt_3FD-THESEUS}, 
the STAR data \cite{STAR:2020dav} are better reproduced within the crossover scenario. 
At 4.5 GeV, the 1PT proton flow turns out to be almost identical to that for the hadronic EoS, 
in spite of that the 1PT trajectory touches the mixed phase of the 1PT transition, 
see Fig. \ref{fig:T-nB_b=6fm}.  
This is because the mixed phase is reached only in the narrow central region of the colliding system and for a short time.

At collision energies above 4.5 GeV, the crossover $v_1$ becomes distinctly preferable 
as compared to hadronic and 1PT results. 
As seen from Fig. \ref{fig:v1-STAR-fxt-4-12GeV}, the hadronic EoS completely fails to reproduce the available 
data \cite{STAR:2014clz} at 7.7 GeV. The 1PT directed flow behaves precisely as it was predicted 
in Refs. \cite{CR99,Br00,St05}. The proton $v_1$ first collapses in the midrapidity, then it demonstrates 
antiflow (i.e. the negative $v_1$ slope in the midrapidity), after that the flow gradually returns 
to the normal flow pattern. This antiflow is the effect of the softest-point region 
that is demonstrated in Fig. \ref{fig:Cs-vs-nb-tph-b6}. 
It is usually claimed that the directed flow is formed at the early stage of the
collision. This is indeed true for the normal component of the directed flow \cite{Br00}. 
Flow and antiflow develop in different rapidity regions \cite{Br00}. 
In the early compression stage of the reaction
the spectators near projectile and target rapidities are deflected by the pressure in the central
hot and dense zone, producing the normal flow. When the expansion of the hot
and dense zone proceeds, the antiflow develops around
midrapidity. In the course of the expansion, the antiflow occupies a broader region around midrapidity
than the normal flow, while the normal flow dominates the region near projectile and target
rapidities.
When the collision energy increases, the system quickly 
passes the corresponding softest-point region due to accumulated expansion inertia. Therefore, 
the influence of that region is essentially reduced. Therefore, the normal flow again becomes dominant 
in the midrapidity. 
Thus, the wiggle ``flow-antiflow-flow'' is a signal of onset of the phase transition.  
As seen from Fig. \ref{fig:v1-STAR-fxt-4-12GeV}, the strong antiflow, that is predicted by the 1PT scenario, 
is not supported by the available data at 7.7 GeV \cite{STAR:2014clz}.

\begin{figure}[!hbt]
\includegraphics[width=.35\textwidth]{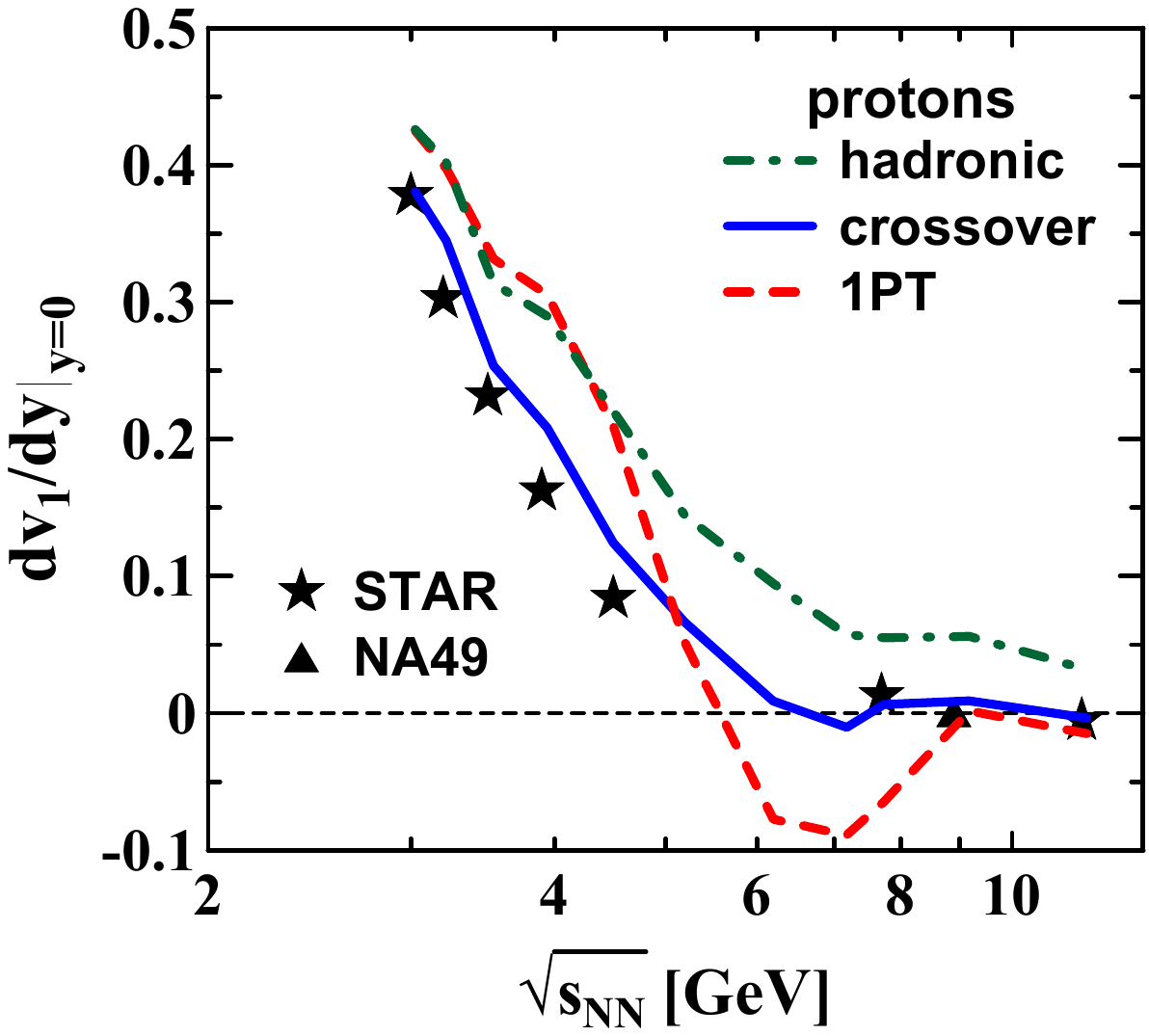}
  \caption{(Color online)
Midrapidity slope of the proton $v_1$ as a function of the collision energy $\sqrt{s_{NN}}$
in semicentral ($b=$ 6 fm) Au+Au collisions. 
Results are calculated within the 3FD model with hadronic, 1PT, and crossover EoSs. 
The  data are from Refs. \cite{STAR:2020dav,STAR:2014clz,STAR:2021yiu,STAR:2025twg} (STAR) 
and \cite{NA49} (NA49). 
}
    \label{fig:dv1dy_ycm_p}
\end{figure}

\begin{figure}[!hbt]
\includegraphics[width=.35\textwidth]{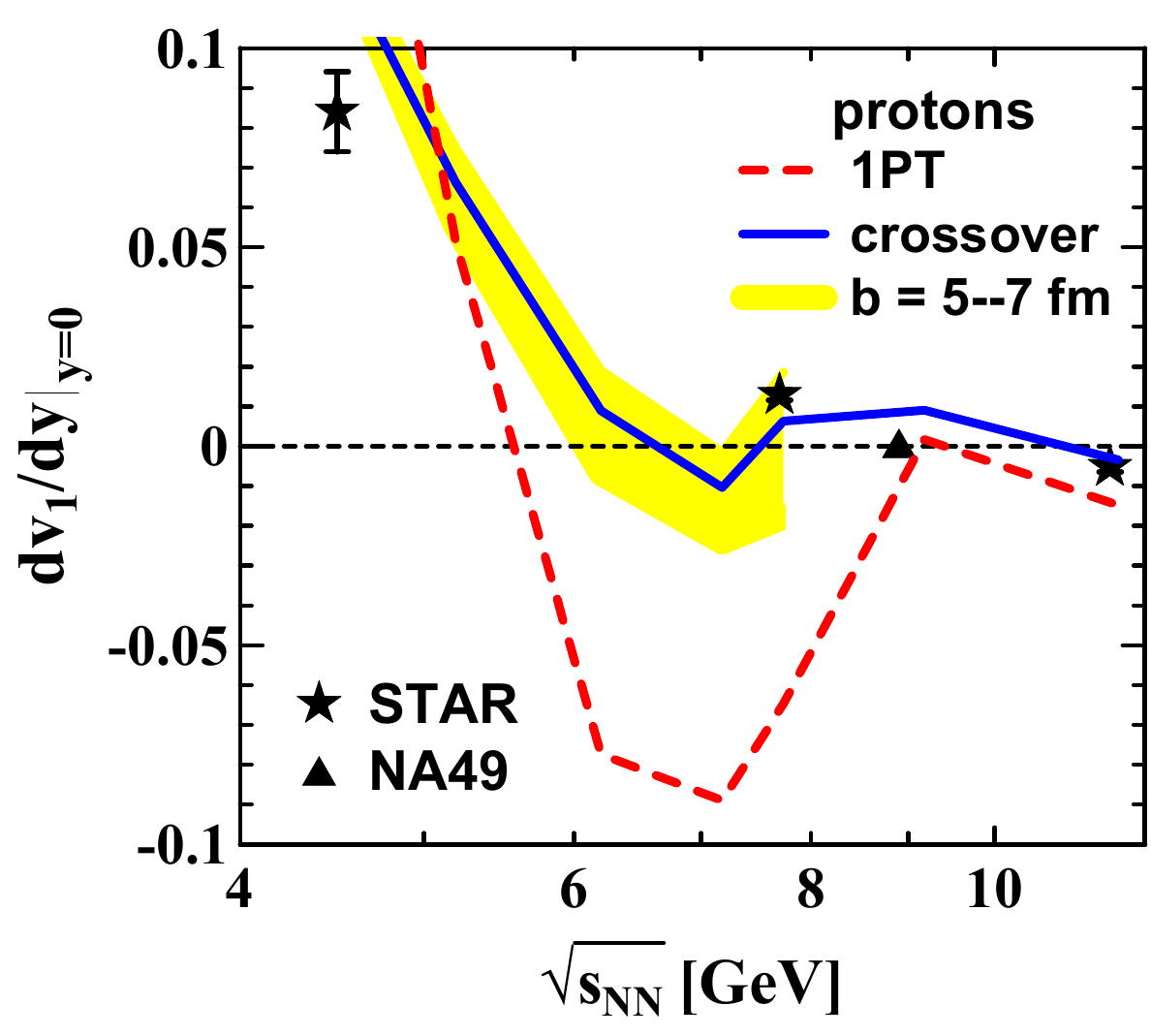}
  \caption{(Color online)
The same as in Fig. \ref{fig:dv1dy_ycm_p} but for the zoomed energy range of 4--12 GeV. 
The area between the curves corresponding to $b=$ 5 and 7 fm is shaded. 
The  data are from Refs. \cite{STAR:2020dav,STAR:2014clz} (STAR) 
and \cite{NA49} (NA49). 
}
    \label{fig:dv1dy_ycm_p-zoomed}
\end{figure}

The crossover $v_1$ predictions well agree with available data \cite{STAR:2020dav,STAR:2014clz}. 
From the first glance, the midrapidity $v_1$ slope monotonously decreases between 4.5 and 7.7 GeV. 
However, a more thorough inspection of this energy region indicates that a weak proton antiflow 
takes place at the energy of 7.2 GeV, see Fig. \ref{fig:v1-STAR-fxt-7,2GeV}. 
\begin{figure}[!hbt]
\includegraphics[width=.35\textwidth]{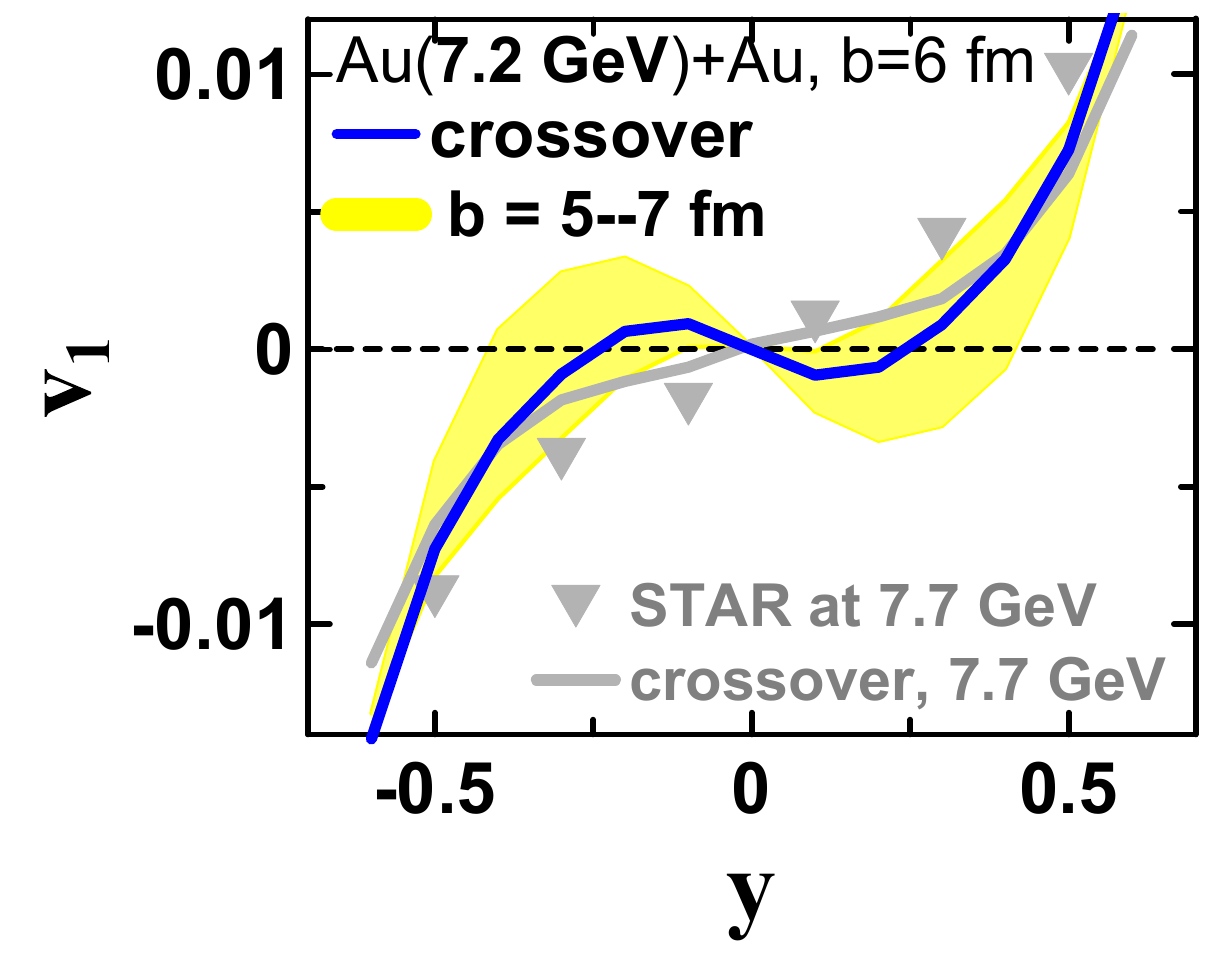}
  \caption{(Color online)
The same as in Fig. \ref{fig:v1-STAR-fxt-3-3,9GeV} in zoomed midrapidity region but at 
$\sqrt{s_{NN}}=$  7.2 GeV and only for crossover EoS. 
The area between additional curves corresponding to $b=$ 5 and 7 fm is shaded. 
STAR data for 7.7 GeV \cite{STAR:2014clz} and the corresponding crossover $v_1$ are also displayed 
in order to visualize the difference between these energies. 
}
    \label{fig:v1-STAR-fxt-7,2GeV}
\end{figure}
In order to visualize the difference between energies 7.2 and 7.7 GeV, 
the STAR data \cite{STAR:2014clz} and the corresponding crossover $v_1$ at 7.7 GeV are also displayed 
in Fig. \ref{fig:v1-STAR-fxt-7,2GeV}. 
To illustrate sensitivity of the antiflow to the collision centrality, 
the $v_1$ results for impact parameters $b=$ 5 and 7 fm are also displayed (shaded area).
This antiflow makes the $v_1$ evolution with the collision energy rise 
to be very similar to the case of the 1PT EoS. To see this similarity more clearly, let us consider the 
excitation function of the midrapidity $v_1$ slope that is presented in Fig. \ref{fig:dv1dy_ycm_p}. 
Both the 1PT and crossover $dv_1/dy$ exhibit dips at 7.2 GeV with negative values of the slope then increase to 
positive values. This is more distinctly seen in Fig. \ref{fig:dv1dy_ycm_p-zoomed}, where 
zoomed energy range of 4--12 GeV is presented. 

Figure \ref{fig:dv1dy_ycm_p-zoomed} demonstrates the slope excitation functions within the 1PT and crossover 
are qualitatively similar, but the amplitude of the wiggle in the crossover scenario is much less than 
that in the strong first-order transition to QGP (1PT EoS). 
The first change of sign followed by minimum in the excitation function of the midrapidity slope of the proton $v_1$ 
results from onset of (1PT or crossover) transition to QGP. The second change of sign
results from correlation of the incomplete baryon stopping and transverse expansion of the system 
that was predicted in Ref. \cite{Snellings:1999bt}. This conclusion is confirmed by the fact that both the 1PT 
and crossover scenarios result in very similar $v_1$ at colision energies above 9 GeV
despite the great difference in the nature of the transition into QGP. This conclusion is also supported by 
findings in Ref. \cite{Du:2022yok}, where it was demonstrated  that 
thorough tune of the baryon stopping allows for good reproduction of the data on the directed flow above 7.7 GeV
\cite{STAR:2014clz}, including the observed  change of sign of the proton $dv_1/dy|_{y=0}$
around 10 GeV collision energy. Authors of Ref. \cite{Nara:2021fuu} also reported that this change of sign 
around 10 GeV results from interplay between baryon stopping and transverse expansion rather than a phase transition. 

The antiflow at 7.2 GeV strongly depends on centrality (impact parameter) of the collision, as seen from 
Figs. \ref{fig:v1-STAR-fxt-7,2GeV} and \ref{fig:dv1dy_ycm_p-zoomed}. While the antiflow transforms into 
inflection point on the rapidity  dependence of the flow at $b=$ 5 fm 
($\approx$10--20\% centrality according to the Glauber calculator \cite{web-docs.gsi.de}), 
see Fig. \ref{fig:v1-STAR-fxt-7,2GeV}, 
it becomes 2.4 times stronger (in terms of the midrapidity $dv_1/dy$, see Fig. \ref{fig:dv1dy_ycm_p-zoomed}) 
at $b=$ 7 fm ($\approx$20--40\% centrality \cite{web-docs.gsi.de}). As has been already mentioned, the proton flow 
is, in general, weakly changed by the afterburner evolution. However, in view of this fragility of the antiflow, 
its sensitivity to the afterburner should be tested.

The available data on the proton $v_1$ exclude any strong phase transition 
while favour a weak phase or crossover transition
to QGP. This conclusion that has been already made earlier in 
Refs. \cite{Konchakovski:2014gda,Ivanov:2014ioa,Ivanov:2016sqy}, 
agrees with that in Refs. \cite{Du:2022yok,Steinheimer:2022gqb}.

%_________________________________________________________________
\section{Summary}
\label{Summary}

The directed flow of various particles provides information on dynamics in various parts 
and at various stages of the colliding system depending on the particle. 
The information on the EoS is not always directly accessible 
because of strong influence of the afterburner stage or insufficient thermalization of the considered probe.  
The proton directed flow gives the most direct information on the EoS 
in heavy-ion collisions at high baryon densities 
because it is minimally affected by the afterburner.

In this paper, new data on the proton directed flow \cite{STAR:2025twg} at energies 3.2, 3.5 and 3.9 GeV were considered. 
Also predictions of the proton directed flow in the energy range between 4.5 and 7.7 GeV were done. 
Calculations were performed within the 3FD  model \cite{Ivanov:2005yw,Ivanov:2013wha}. 
The calculations with the crossover EoS \cite{Toneev06}  well reproduced the proton directed flow 
\cite{STAR:2020dav,STAR:2021yiu}
at collision energies both below 4.5 GeV \cite{Ivanov:2024gkn}, 
including the energies 3.2--3.9 GeV considered in this paper, 
and above 7.7 GeV  \cite{Konchakovski:2014gda,Ivanov:2014ioa,Ivanov:2016sqy}. 
In addition, the crossover EoS is favorable for bulk observables \cite{Ivanov:2013yqa} in the whole energy range of interest. 
Therefore, the crossover predictions for the proton directed flow 
at energies from 4.5 to 7.7 GeV can be considered as reliable predictions. Predictions with two other 
(hadronic and 1PT) EoSs  were also presented to illustrate sensitivity of the proton directed flow to the EoS.

It is predicted that the proton flow evolves non-monotonously in the energy range from 4.5 to 7.7 GeV. 
At the energy of 7.2 GeV it exhibits antiflow (i.e. negative slope of 
$v_1(y)$) in the midrapidity. At 7.7 GeV, the flow returns to the normal pattern in accordance with the 
STAR data \cite{STAR:2014clz} and then again exhibits antiflow at 11.5 GeV 
also in agreement of the STAR data \cite{STAR:2014clz}.

Thus, the $v_1$ slope excitation functions within the 1PT and crossover scenarios turned out to be 
qualitatively similar, but the amplitude of the wiggle in the crossover scenario being much less than 
that in the strong first-order transition to QGP. 
The first change of sign followed by minimum in the excitation function of the midrapidity slope of the proton $v_1$ 
results from onset of (1PT or crossover) transition to QGP. The second change of sign
results from correlation of the incomplete baryon stopping and transverse expansion of the system 
that was predicted in Ref. \cite{Snellings:1999bt}.

The available data on the proton $v_1$ exclude any strong phase transition 
while favour a weak phase or crossover transition to QGP. 
Experimental observation of the wiggle in excitation function of the  
midrapidity slope of the proton directed flow at collision energy of 7.2 GeV 
may confirm this conclusion.

% ________________________________________________________________
\begin{acknowledgments}

Fruitful discussions with  D.N. Voskresensky are gratefully acknowledged.
This work was carried out using computing resources of the federal collective usage center ``Complex for simulation and data processing for mega-science facilities'' at NRC "Kurchatov Institute" \cite{ckp.nrcki.ru}.
%and computing resources of the supercomputer "Govorun" at JINR \cite{govorun}. 

\end{acknowledgments}
% ________________________________________________________________

%_________________________________________________________________
\section*{DATA AVAILABILITY}

Tabulated 1PT and crossover EoSs that were used in the present simulations are publicly available 
on GitHub \cite{github.com/marinakozh}. 
The data that support the findings of this article are openly available.

% ________________________________________________________________

%%%%%%%%%%%%%%%%%%%%%%%%%%%%%%%%%%%%%%%%%%%%%%%%%%%%%
\end{document}